\documentclass[prl,twocolumn,showpacs,preprintnumbers,amsmath,amssymb]{revtex4}

\usepackage{graphicx}
\usepackage{dcolumn}
\usepackage{bm}

\begin{document}
\preprint{prepared for Arxiv Preprint}

\title{Anomalous Expansion of the Copper-Apical Oxygen Distance in Superconducting La$_{2}$CuO$_{4}$ $-$ La$_{1.55}$Sr$_{0.45}$CuO$_{4}$ Bilayers}

\author{Hua Zhou$^{1}$}
\author{Yizhak Yacoby$^{2}$}
\author{Ron Pindak$^{1}$}\email{pindak@bnl.gov}
\author{Vladimir Butko$^{1}$}
\author{Gennady Logvenov$^{1}$}
\author{Ivan Bozovic$^{1}$}
\affiliation{$^{1}$Brookhaven National Laboratory, Upton, NY 11973 USA\\
$^{2}$Racah Institute of Physics, Hebrew University, Jerusalem 91904, Israel}
\date{\today}

\begin{abstract}
We have introduced an improved X-ray phase-retrieval method with unprecedented speed of convergence and precision, and used it to determine with sub-{\AA}ngstrom resolution the complete atomic structure of an ultrathin superconducting bilayer film, composed of La$_{1.55}$Sr$_{0.45}$CuO$_{4}$ and La$_{2}$CuO$_{4}$ neither of which is superconducting by itself. The results show that phase-retrieval diffraction techniques enable accurate measurement of structural modifications in near-surface layers, which may be critically important for elucidation of surface-sensitive experiments. Specifically we find that close to the sample surface the unit cell size remains constant while the copper-apical oxygen distance shows a dramatic increase, by as much as 0.45 {\AA}. The apical oxygen displacement is known to have a profound effect on the superconducting transition temperature.
\end{abstract}

\maketitle

The exciting discovery of interface superconductivity in complex oxides \cite{Reyren,Ivan1,Ivan6,Ueno,Yuli} has triggered intense debate about its origin and the possibility to enhance T$_{c}$ even further \cite{Kivelson,Yunoki,Okamoto,Berg}. The interfacial enhancement \cite{Ivan6} of the superconducting critical temperature (T$_{c}$) is influenced by the crystal structure. The Z-axis lattice constant (c$_{0}$) varies significantly among La$_{2}$CuO$_{4}$ (LCO), La$_{1.55}$Sr$_{0.45}$CuO$_{4}$ (LSCO), and bilayer LSCO/LCO films, depending even on the deposition sequence, and it affects superconductivity: T$_{c}$ scales with c$_{0}$ almost perfectly linearly \cite{Butko}. The reason for this is not understood at present, but notice that in (La,Sr)$_{2}$CuO$_{4}$ the change in c$_{0}$ goes together with the change in c$_{A}$, the distance between copper and the nearest apical oxygen, which some believe to play a key role in the high temperature superconductivity (HTS) phenomenon \cite{Fernandes,Ohta,Feiner,Lubrittoy,Pavarini,Slezak}. In any case, it is certain that (i) from one cuprate to another, c$_{A}$ varies more than any other bond length, and (ii) at least in simple cuprates with a single CuO$_{2}$ layer in the unit cell it correlates with the maximal T$_{c}$ - the longer c$_{A}$, the higher T$_{c}$. At least, the first fact can be understood: c$_{A}$ is 'soft' because apical oxygen has no hard contact with the nearest copper ion; rather, it "levitates" on the electrostatic potential - a structural feature peculiar to certain layered oxides with alternating ionic planes of opposite charge \cite{Ivan5}. This makes apical oxygen prone to very large displacements - e.g., in HgBa$_{2}$CuO$_{6}$ one finds c$_{A}$ $\approx$ 2.8 {\AA}, longer by 0.9 {\AA} than the in-plane Cu-O bond; coincidentally, this compound has the highest T$_{c}$ = 97 K among all single-CuO$_{2}$-layer cuprates \cite{Wagner}.

It is thus important to find out what happens to the apical oxygen in LSCO-LCO bilayers; however, standard X-ray diffraction (XRD) is not well suited for this $-$ one needs to "get inside the unit cell" and look for individual atomic displacements. For this, the most suitable technique is the Coherent Bragg Rod Analysis (COBRA) method \cite{COBRA1,COBRA2,COBRA3,COBRA4,COBRA5,COBRA6}. However, COBRA is most effective for films that are just a few unit cells (UCs) thick, and in the case of HTS compounds fabrication of ultrathin films with bulk properties has been proven to be extremely challenging. Fortunately, we had a technical solution at hand $-$ a unique atomic layer-by-layer molecular beam epitaxy (ALL-MBE) system with proven capability of fabricating ultrathin HTS layers \cite{ALLMBE,Ivan3}.

For this study we have synthesized by ALL-MBE a number of (n $\times$ LSCO + m $\times$ LCO) bilayers, where (n,m) determine the thickness of the respective layers expressed as the number of UCs. In this paper, we show the COBRA results for two of these, (2.5,2.5) and (2,3). The films were deposited at T = 650$^{\circ}$C and p = 9 $\times$ 10$^{-6}$ Torr of ozone and subsequently cooled down under high vacuum to drive out all the interstitial oxygen. We used 10 $\times$ 10 $\times$ 1 mm $^{3}$ single-crystal LaSrAlO$_{4}$ (LSAO) substrates polished with the large surface perpendicular to the (001) direction. The substrate lattice constants are a$_{0}$ = b$_{0}$ = 3.755 {\AA}, c$_{0}$ = 12.56 {\AA}; the films are pseudomorphic with LSAO and under compressive strain. The crystal structure of LCO is illustrated in Fig. 1a. Atomic force microscopy scans over a large (10 $\times$ 10  $\mu$m$^{2}$) area showed root-mean-square surface roughness of 0.25 nm in the (2,3) and 0.11 nm in the (2.5,2.5) bilayer sample; this is significantly less than the 1 UC step height which in LSCO is 1.32 nm. Magnetic susceptibility was measured via two-coil mutual inductance technique and revealed sharp superconducting transitions at T$_{c}$ = 34 K in the (2,3), and T$_{c}$ = 36 K in the (2.5,2.5) bilayer, significantly higher than the values reported for (n,m) bilayers in Ref. [3], which is remarkable given that these films are only 5 UC thick. This was also confirmed by measuring the electric resistance (see Fig. 1b) after the X-ray scattering experiments were completed and gold pads were evaporated to enable four-point-contact measurements.

\begin{figure}
  \centering
  \includegraphics[scale=0.36]{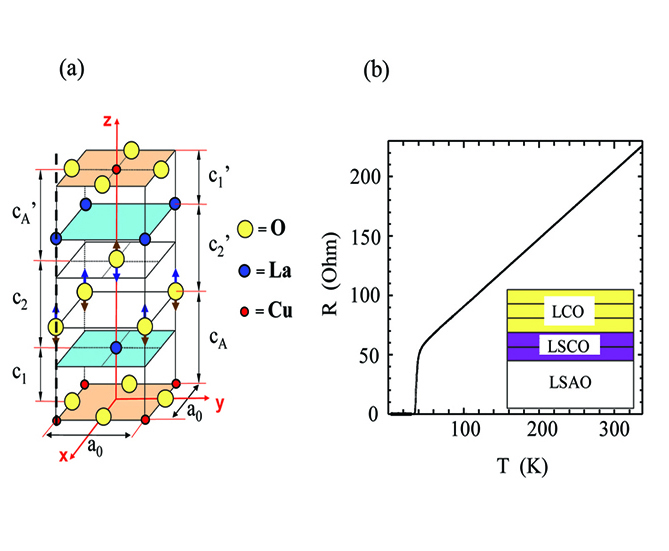}
  \caption{(Color online) A simplified structure model (one-half the crystallographic unit cell) of La$_{2}$CuO$_{4}$ and the transport property for a bilayer film. (a) At room temperature, the structure is tetragonal and the space group is I4/mmm. Noted that the La(Sr)-O layers are strongly corrugated, exaggerated in this sketch for clarity. (b) The electric resistance of the (2.5,2.5) bilayer, measured by the four-point-contact technique, as a function of temperature. Inset: a schematic of the bilayer on a LSAO substrate.}
\end{figure}

The atomic structure of the LSCO/LCO bilayer film was investigated at beamline ID-33 of the Advanced Photon Source by measuring the diffraction intensities along the substrate-defined Bragg rods. The sample and a PILATUS 100k photon-counting pixel detector \cite{Pilatus} were mounted on a six-circle goniometer in Kappa geometry. The experimental set-up and procedures were described in detail in previous works \cite{COBRA2,COBRA4}. Ten symmetry inequivalent Bragg rods were recorded with a maximum value for the vertical reciprocal space coordinate of L$_{max}$ = 10.5 r.l.u. (reciprocal lattice units) and a sampling density of 50 points per r.l.u.. The X-ray flux after the Si (1,1,1) monochromator crystal was 3$\times10^{12}$ photons/sec at a wavelength of $\lambda$ = 0.8266 {\AA}. For all Bragg rod measurements, except for the (0,0,L) rod, the angle of incidence had a fixed value of 3.5$^{\circ}$. The X-ray beam was focused to 0.1 mm (V) $\times$ 0.2 mm (H), resulting in a 2 mm long X-ray footprint. The background and diffuse X-ray scattering contribution were removed efficiently and accurately using the PILATUS detector images. The final results were then normalized by taking into account the beam polarization and Lorentz factors. The results were subsequently analyzed using the COBRA method \cite{COBRA1,COBRA3,COBRA4,COBRA5,COBRA6}. In general, COBRA uses the measured diffraction intensities and the fact that the complex structure factors (CSFs) vary continuously along the substrate-defined Bragg rods to determine the diffraction phases and the CSFs. The CSFs are then Fourier transformed into real space to obtain the 3-dimensional electron density of the film and the substrate with sub-{\AA}ngstrom resolution.

\begin{figure}
  \centering
  \includegraphics[scale=0.36]{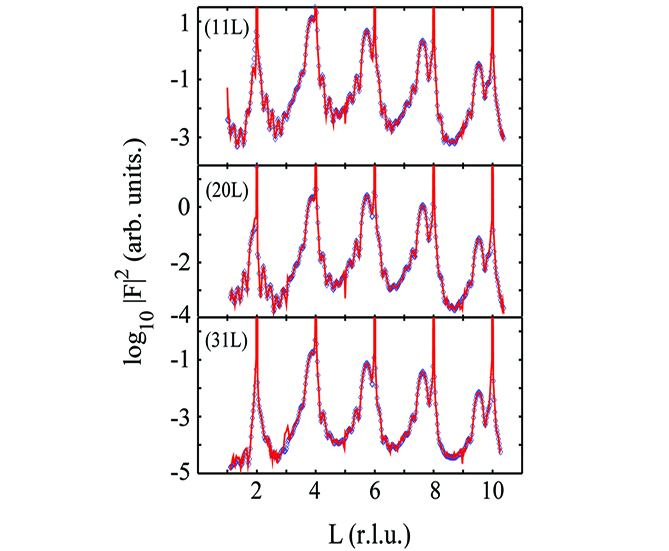}
  \caption{(Color online) Representative Bragg rods of LSCO/LCO system (open diamond) and calculated diffraction intensity obtained from COBRA-determined electron density (solid line).}
\end{figure}

The experimental data of three representative Bragg rods are shown in Fig. 2(a). Notice that the diffraction intensity along the Bragg rods (excluding the Bragg peaks) vary over more than 4 orders of magnitude with excellent signal-to-noise ratio. The reference structures chosen as the starting point for the COBRA analysis were the bulk LSAO structure and the tetragonal LSCO/LCO bilayer with the nominal layer atomic positions. In our numerical simulations, the topmost 4 UCs of the substrate were allowed to deform, however the resulting deformations turned out to be very small. The COBRA method uses the approximation that at two adjacent points along the Bragg rod the change in CSFs contributed by the unknown part of electron density is negligible compared to the change in CSFs contributed from the reference structure \cite{COBRA2}. The use of this approximation allows COBRA to converge very quickly to approximately the right solution but not to the exact one. To overcome this limitation we further refined the CSFs using the Difference-Map algorithm introduced by Elser \cite{Elser} and recently applied to thin films \cite{Bjorck}. Using the COBRA solution as the starting point for the Difference-Map algorithm and using a proper filter program that takes advantage of the fact that the CSFs vary continuously along the Bragg rods, the Difference-Map algorithm converges after about 20 iterations; the convergence accelerates by about two orders of magnitude. As seen in Fig. 2, the final calculated and measured intensities are in very good agreement. Similar agreement was found for all other Bragg rods and the overall X-ray reliability factor \begin{math}R = \frac{\Sigma ||F_{0}|-|F_{c}||}{\Sigma |F_{0}|} = 0.02\end{math}; here, F$_{0}$ and F$_{c}$ are the observed and the calculated diffraction amplitudes, respectively. To the best of our knowledge, so far there has been only one attempt to determine the structure of a thin film using the Difference Map method \cite{Bjorck}. In that study, the atomicity constraint was imposed and over 2,000 iterations were needed to achieve convergence. Our analysis shows that the combined COBRA/Difference Map method combines the best features of both methods and ensures rapid convergence to the correct solution without the need to use the atomicity constraint.

\begin{figure}
  \centering
  \includegraphics[scale=0.38]{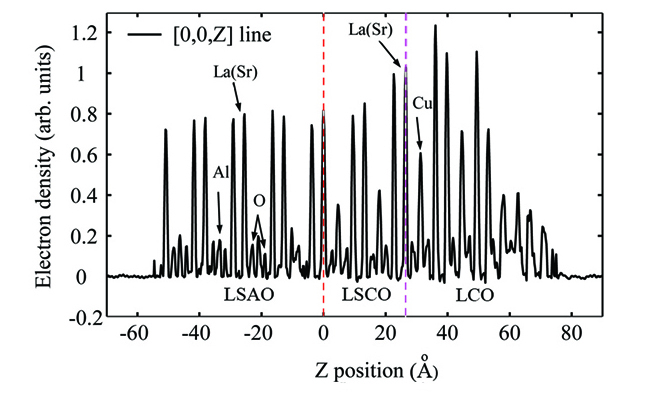}
  \caption{(Color online) The electron density variation, determined by COBRA, along the [0, 0, Z] column of atoms as indicated by the dashed line in Figure 1a.  Note that the topmost four unit cells of the substrate are included in the structure refinement. The left and right dashed lines represent the nominal LSAO/LSCO and LSCO/LCO interfaces, respectively.}
\end{figure}

The CSFs obtained have been Fourier transformed into real space yielding the 3D electron density (ED). As an example, we show in Fig. 3 the ED of a (2,3) bilayer sample along the [0,0,Z] line that goes through the La(Sr), O, and Cu(Al) atoms. Two points should be stressed. First, as seen the ED has almost no negative parts. Together with the excellent agreement between the calculated and measured diffraction intensities, this suggests that the ED is very close to the correct one. Second, all the atoms including the oxygens can be clearly identified and their positions determined with sub-{\AA}ngstrom resolution. The small ED intensity fluctuation below -53 {\AA} provides a measure of the inaccuracy in the ED and as seen it is small even compared to the oxygen ED.

The atomic positions in the Z direction were accurately determined by fitting a Gaussian to each peak. We determined the size of the unit cell in the Z direction by measuring the distance between consecutive pairs of La(Sr) and Cu atoms. The results are shown in Fig. 4 inset. Each point corresponds to an average of 4 La-La distances and 2 Cu-Cu distances. The measured lattice constant of the bilayer film is 13.304 $\pm$ 0.016 {\AA} and is larger by 0.148 {\AA} than that of the bulk LCO (c$_{0}$ = 13.156 {\AA}).

\begin{figure}
  \centering
  \includegraphics[scale=0.38]{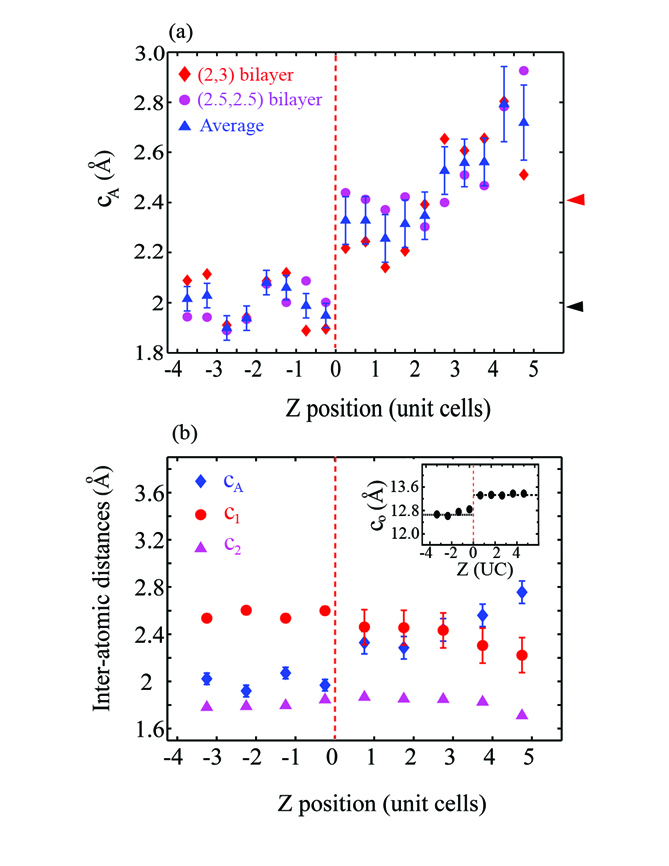}
  \caption{(Color online) Evolution of the inter-atomic distances. (a) The measured Cu(Al)$-$ apical O distance, c$_{A}$, varies as a function of the nominal position of Cu(Al) atoms inside the refined structure. The data from two representative bilayer samples and the average over the two are presented. The lower and upper arrows represent the bulk values of c$_{A}$ for LSAO and for LCO, respectively. (b) The comparison of c$_{A}$, c$_{1}$, and c$_{2}$, averaged for each unit cell, as a function of Z position. Inset: the lattice constant c$_{0}$ as a function of Z. The dotted line represents the bulk LSAO value. The horizontal dashed line is the average value of c$_{0}$ in bilayers extracted from the electron density, as described in the text. In both (a), (b) and the inset, the vertical dashed lines represent the nominal LSAO/LSCO interfaces, respectively.}
\end{figure}

While the changes observed in the unit cell sizes are as expected from the strain and the elastic parameters \cite{Poisson}, the variations in Cu-apical O and La-apical O distances are quite unexpected. The distances c$_{A}$,c$_{1}$ and c$_{2}$ are defined in Fig. 1a; the distances labeled c$_{A}^{'}$,c$_{1}^{'}$ and c$_{2}^{'}$ would be their symmetry equivalents in bulk samples, but in thin films they could differ in principle. For our samples, the measured values for the primed and unprimed distances were in fact equal within the experimental error, except at the LSAO/LSCO interface. The measured values averaged over c$_{A}$ and c$_{A}^{'}$ are shown in Fig. 4a. The diamond and circular dots represent the distances measured in (2,3) and (2.5,2.5) bilayers, respectively. The triangular dots are averages over the two samples. Every pair of triangular dots corresponds to one UC. The dashed vertical lines represent the nominal LSAO/LSCO interfaces. The arrows on the right indicate c$_{A}$ as measured in the bulk samples. The results show that, within the experimental error, the values of c$_{A}$ in the substrate are equal to those in the bulk but they are very different in the film. In both LSAO and LCO bulk crystals \cite{Ivan5}, c$_{A}$ is equal to 2.41 {\AA}. In the metal layer closest to the substrate c$_{A}$ = 2.3 {\AA}, and it then rises steadily all the way to c$_{A}$ = 2.75 {\AA} $-$ a change of 0.45 {\AA}.

In Fig. 4b we display c$_{A}$ as well as the La-apical O distance, c$_{2}$, and the La-CuO$_{2}$ plane distance, c$_{1}$. Each point represents an average over the two bulk-symmetry-equivalent distances and over the two measured samples. Notice that c$_{1}$ changes by less than 0.1 {\AA}. On the other hand, c$_{A}$ increases by about 0.45 {\AA}, while c$_{2}$ decreases by about 0.25 {\AA}. Close to the film surface, the apical oxygen atoms are displaced away from the nearest Cu atoms. The La atoms are displaced towards the closest CuO$_{2}$ plane, but by a smaller amount, while the separation between two adjacent CuO$_{2}$ planes remains constant.

According to Ref. 15, c$_{A}$ = 2.7 {\AA} should correspond to a T$_{c}$ of $~$80 K at the optimum doping. However, from Ref. 31 we know that the hole density drops sharply on the I side of the interface and the screening length is equal to 6 $\pm$ 2 {\AA}. This implies that on the I side and next to the M-I interface only one or two CuO$_{2}$ layers are doped via carrier accumulation while the others remain insulating. Thus, unfortunately, we have a mismatch: in the optimally-doped LCO layer, c$_{A}$ is close to its standard (bulk) value, while it is greatly elongated only in insulating layers. It is tempting to speculate that one could create LSCO-based samples with T$_{c}$ much higher than 36 K, perhaps as high as 80-90 K, if only one could achieve c$_{A}$ elongation and optimal doping in the same LCO layer. An obvious avenue for further research is to try making I layers even thinner, thus bringing the interface superconductivity closer to the film surface. Another is to try engineering more sophisticated hetero-structures and superlattices combining LCO with other metallic oxides (nickelates, zincates, etc.).

In summary, we have used ALL-MBE to synthesize precise ultrathin bilayers using metallic but non-superconducting LSCO and insulating LCO blocks, and observed interface superconductivity with T$_{c}$ = 34-36 K, significantly higher than before. We have used synchrotron X-ray diffraction and the combined COBRA/Difference-Map phase-retrieval method to determine accurately the atomic structure and found the unit cell size to be constant despite dramatic atomic displacements within the cell. In particular, near the surface the Cu$-$apical O increases greatly, by as much as 0.45 {\AA}, while it is known that variations in the apical oxygen position strongly affect T$_{c}$. We conclude that in cuprates the crystal structure can be modified in near-surface layers, and in such a way that superconductivity properties can be dramatically altered. This result amplifies the importance of high quality surface structure determination in conjunction with surface sensitive probes of electronic states such as scanning tunneling microscopy or angle-resolved photoemission spectroscopy.

This work was supported by the U.S. Department of Energy, Office of Science, Office of Basic Energy Sciences, Project MA-509-MACA, Contract No. DE-AC02-98CH10886, and use of the Advanced Photon Source under Contract No. DE-AC02-06CH11357. One of us (Y. Yacoby) would like to acknowledge with thanks fruitful discussions with M. Bj\"{o}rck.


\end{document}